\documentclass[titlepage,12pt,subeqn]{utarticle}
\usepackage{amsmath,amssymb,amscd}
\usepackage{color}
\usepackage{latexsym}
\usepackage{ifthen}
\usepackage[]{graphicx}
\usepackage[]{hyperref} 
\usepackage{bbm}

\numberwithin{equation}{section}

\newcommand{\be}{\begin{equation}}
\newcommand{\ee}{\end{equation}}
\newcommand{\bea}{\begin{eqnarray}}
\newcommand{\eea}{\end{eqnarray}}
\newcommand{\al}{\alpha}
\renewcommand{\d}{\delta}
\newcommand{\e}{\epsilon}

\newcommand{\g}{\gamma}

\newcommand{\La}{\Lambda}
\newcommand{\la}{\lambda}
\newcommand{\m}{\mu}
\newcommand{\n}{\nu}

\newcommand{\s}{\sigma}
\renewcommand{\t}{\theta}

\newcommand{\hlf}{\frac{1}{2}}

\newcommand{\p}{\partial}

\newcommand{\lp}{\left(}
\newcommand{\rp}{\right)}
\newcommand{\ls}{\left[}
\newcommand{\rs}{\right]}
\newcommand{\ol}{\overline}
\newcommand{\wt}{\widetilde}
\newcommand{\Lra}{\Longrightarrow}

\begin{document}
\preprint{
UTTG-07-08 \\
}
\title{General F-Term Gauge Mediation}

\author{ {\sc Jacques Distler} and {\sc Daniel Robbins}
     \oneaddress{
      Theory Group, Department of Physics,\\
      University of Texas at Austin,\\
      Austin, TX 78712, USA \\
      {~}\\
      \email{distler@golem.ph.utexas.edu}\\
      \email{robbins@zippy.ph.utexas.edu}\\
      }
}

\date{July 12, 2008}

\Abstract{In a model-independent formalism of gauge mediation, Meade, Seiberg, and Shih have shown that hidden sector effects are captured by two-point correlation functions of the gauge current superfields and that, generically, many of the characteristic features of gauge mediated SUSY breaking do not survive.  We review the general story, particularly the way that the correlators enter the low-energy effective action and give rise to soft-breaking terms. We then specialize to the case where there is a small parameter, $F/m^2$, where $m$ is the mass scale characterizing the hidden sector, and $F$ is the strength of the SUSY breaking. To leading nontrivial order in this small parameter, we show that many of the classic predictions of gauge mediated SUSY breaking are recovered.}

\maketitle

\section{Introduction}
Gauge mediation is an attractive scenario for supersymmetry breaking because it naturally leads to flavour-blind sfermion masses, skirting the would-be problem of Flavour-Changing Neutral Currents. In the context of string theory, it is also attractive because (unlike, say, gravity mediation) it can be discussed in a decoupling limit, in which one sends the 4-dimensional $M_{\text{Pl}}\to\infty$. There is, of course, a vast literature on gauge mediation (see \cite{Giudice:1998bp}) for a review), but the plethora of models share  several characteristic features, which might be called ``predictions'' of the gauge mediation scenario

\begin{itemize}
\item The LSP is the gravitino.
\item The NLSP is typically the bino or the stau.
\item Gaugino mass unification: the ratio of gaugino masses, $\tfrac{m_{r}}{m_{r'}} = \tfrac{g^{2}_{r}}{g^{2}_{r'}}$, $r,r'=1,2,3$.
\item Small A-terms.
\end{itemize}

Several months ago, Meade {\it et al} \cite{Meade:2008wd} presented a model-independent framework for models of gauge mediation. Unfortunately, in the context of this general framework, most of these ``characteristic'' features of gauge mediated SUSY breaking do not appear to emerge. Of course, the gravitino is still the LSP. But the only other predictions were a pair of sum rules on the sfermion mass matrix (that can also be independently derived~\cite{Cohen:2006qc}),
\begin{equation}
\begin{aligned}
  Tr Y m_{\tilde{f}}^{2} &=0\\
  Tr (B-L) m_{\tilde{f}}^{2} &=0\\
\end{aligned}
\end{equation}
This is rather disappointing as those features which might be seen as a smoking gun for gauge mediation, if seen at the LHC (or, conversely, if not seen, would be viewed as ruling out gauge mediation) do not seem to be generic features of gauge mediation after all.

The reason, however, is easy to understand. In their formulation, the hidden sector is characterized by a single mass scale, $m$. And at that mass scale, SUSY-breaking effects are assumed to be $O(1)$. By contrast, most explicit models have a small parameter, $|F/m^{2}|\ll 1$, where $F$ is the strength of the supersymmetry-breaking.  We will see, in this paper, that, to leading nontrivial order in this small parameter, most of the classic predictions of gauge mediation are recovered.

Typically, in gauge mediated SUSY-breaking, the gaugino masses are generated at one loop, with messenger fields running in the loop, and sfermion masses are generated at two-loops.  In~\cite{Giudice:1997ni} (see also the review~\cite{Giudice:1998bp}), Giudice and Rattazzi observed that both these effects as well as the soft-breaking A-terms, could be obtained from consideration of the one-loop RG equations.  Those authors, along with their collaborators, continued to explore these connections, using considerations of analyticity in superspace~\cite{ArkaniHamed:1998kj}, an approach that is quite similar to that which we follow in this work.

Our approach will be to use the general gauge mediation formalism of~\cite{Meade:2008wd} and consider the relations between soft-breaking terms in the effective action and the two-point correlation functions of the hidden sector.  Next, by considering all possible terms in the effective action which are quadratic in the MSSM gauge fields and which are consistent with our assumptions, we will derive functional relations between the two-point correlators which will allow us to recover the relations between the soft terms and $\beta$-functions which were found in the works above (see also  \cite{Hisano:1997ua}).

We can also use this approach to incorporate D-term SUSY-breaking with a small parameter.  In this case the supersymmetry breaking parameter is an auxiliary VEV for a hidden sector $U(1)$ gauge group, and it again will appear as a spurion in the low-energy effective action.

Anomaly mediation \cite{Giudice:1998xp},\cite{Randall:1998uk} shares many features with the formalism of~\cite{Meade:2008wd}. In particular, the anomaly-mediated contributions to the gaugino and sfermion masses (for a recent discussion, and further references, see \cite{deAlwis:2008aq}) vanishes in the limit of vanishing Standard Model gauge couplings. Thus, many of the present considerations should apply in that context as well.

Another recent work which also used the effective action to learn about the hidden sector correlators in various models is~\cite{Ooguri:2008ez}.

The plan of this paper is as follows.  In section \S\ref{Computations}, we will examine the approach of \cite{Meade:2008wd} from a superspace perspective and recover their formalism in which the effects of the hidden sector are encapsulated in a set of current-current two-point functions.  Section \S\ref{GeneralConnection} discusses how the two-point functions enter into the low-energy effective action and in particular into the computations of masses for gauginos and sfermions.  Section \S\ref{SingleSpurionResults} will summarize our main results.  Section \S\ref{HigherDerivativeTerms} considers possible effective actions which are consistent with our assumptions at arbitrary order in a derivative expansion, and shows how our results follow, including a discussion of the scalar trilinear terms (A-terms). In \S\ref{multispurion}, we comment briefly on the multispurion case.  Finally in \S\ref{DTerms} we apply our technique to D-term SUSY-breaking and find the lowest order contribution to the soft-breaking terms. Details are relegated to an appendix.

\section{Computations}
\label{Computations}

\subsection{Setup}

In the formalism of Meade {\it et al} \cite{Meade:2008wd}, there is a hidden sector, $S$, in which supersymmetry is broken. This sector has a global symmetry group, $G$, and we weakly gauge an $SU(3)\times SU(2)\times U(1)$ subgroup of $G$. When the Standard Model gauge couplings are set to zero, the hidden sector decouples from the MSSM sector, and the latter is supersymmetric.

The currents which couple to the Standard Model gauge fields are part of a real linear supermultiplet,  $\mathcal{J}(p,\t,\ol{\t})$, satisfying\footnote{Throughout we use the conventions of \cite{Wess:1992cp}, so for example the superspace covariant derivative is $D_\al=\frac{\p}{\p\t^\al}+p_\m\s^\m_{\al\dot\al}\ol{\t}^{\dot\al}$.} \begin{equation}
D^2\mathcal{J}=\ol{D}^2\mathcal{J}=0
\end{equation}

When we integrate out the sector $S$, we obtain corrections to the effective Lagrangian describing the MSSM sector. We will mostly be interested in the corrections which are quadratic in the Standard Model gauge multiplets, which is to say that we will be interested in
the two point function of $\mathcal{J}(p,\t,\ol{\t})$. This must have the form\footnote{We will assume an unbroken $\mathbb{Z}_2$ symmetry, which forbids a VEV for $\mathcal{J}$.}
\begin{equation}
\left\langle\mathcal{J}(p,\t,\ol{\t})\mathcal{J}(p',\t',\ol{\t}')\right\rangle=\lp 2\pi\rp^4\d^4\lp p+p'\rp\, I(p,\t,\ol{\t},\t',\ol{\t}'),
\end{equation}
where $I$ is a real function which should satisfy
\begin{equation}
\label{InterchangeSymmetry}
I(p,\t,\ol{\t},\t',\ol{\t}')=I(-p,\t',\ol{\t}',\t,\ol{\t}),
\end{equation}
and also
\begin{equation}
D^2I=\ol{D}^2I=\lp D'\rp^2I=\lp\ol{D}'\rp^2I=0.
\end{equation}
Note that the latter two conditions will be automatic once we impose the symmetry property (\ref{InterchangeSymmetry}).  

If we are in a theory with unbroken supersymmetry, then this two-point function should also be invariant under the supersymmetry transformations, which leads to the condition that
\begin{equation}\label{susycond}
\lp Q_\al+Q_\al'\rp I=\lp\ol{Q}_{\dot\al}+\ol{Q}_{\dot\al}'\rp I=0.
\end{equation}

Imposing all of these conditions determines $I$ in terms of a single scalar function\footnote{We would like to thank Martin Rocek for suggesting the second expression in \eqref{susyresult}. The operator $\tfrac{1}{8}D^\alpha \overline{D}^2 D_\alpha$ is the projector onto linear supermultiplets and $(\theta-\theta')^2(\overline{\theta}-\overline{\theta}')^2=\delta^4(\theta-\theta')$ ensures the supersymmetry of the result.} of $p^{2}$.
\begin{equation}\label{susyresult}
\begin{split}
  I &= e^{(\theta'\sigma^{\mu}\overline{\theta}-\theta\sigma^{\mu} \overline{\theta}')p_{\mu}+ \frac{1}{4} (\theta-\theta')^{2}(\overline{\theta}-\overline{\theta}')^{2}p^{2}} C_{\text{SUSY}}(p^{2})\\
    &= \tfrac{1}{8} D^\alpha \overline{D}^2 D_\alpha (\theta-\theta')^2(\overline{\theta}-\overline{\theta}')^2 C_{\text{SUSY}}(p^{2}) 
\end{split}
\end{equation}

The two point function is singular at short distances\footnote{The position-space superfield, $\mathcal{J}(x,\theta,\overline{\theta})$ has mass dimension 2 (since the conserved current, $j_\mu$ has mass dimension 3). The leading singularity at short distances, thus, $\sim 1/|x|^4$, whose Fourier transform is a logarithm.}. One must introduce a UV cutoff to define $C_{\text{SUSY}}(p^2)$. We can write
\begin{equation}
  C_{\text{SUSY}} = c \log(\Lambda^2/m^2) + f(p^2/m^2)
\end{equation}
where
\begin{equation}
  f(y) \sim \begin{cases} -c\log(y) + \text{finite},&x\to\infty\\
  \text{analytic in}\, x,& x\to 0\end{cases}
\end{equation}
and $c$ is a constant.

As we will see more explicitly below, the constants $c^{(r)}$ (one for each factor in the Standard Model gauge group) govern the dependence of the gauge kinetic terms on the cutoff scale $\Lambda$, and hence are related to the shifts in the $\beta$-functions upon integrating out the the hidden sector.
\begin{equation}
\Delta b^{(r)}=b^{(r)}_{\text{low}}-b^{(r)}_{\text{high}}=16 \pi^2 c^{(r)}.
\end{equation}
where $b^{(r)}$ are the usual beta-function coefficients,
\begin{equation}
\beta(g_r(\m))=\frac{d}{d(\ln\m)}g_r(\m)=-\frac{g_r^3(\m)}{16\pi^2}b^{(r)}(\m),
\end{equation}
and where high and low refer to the theories above and below the messenger scale $m$, respectively.
If we want the gauge couplings to unify, then we should arrange that the shifts in the $\beta$-function coefficients have a universal value, $c^{(r)}\equiv c$, independent of the gauge group. This can be achieved, for instance, by arranging the global symmetry group, $G$, to contain $SU(5)$ as a subgroup.

When SUSY is broken, we should relax the condition \eqref{susycond}, and write the most general solution compatible with the remaining conditions. In agreement with Meade {\it et al} \cite{Meade:2008wd}, we will find four independent contributions to $I$, once SUSY is broken.

\subsection{The general solution}

Note that we can define separate R-charges here for the primed and unprimed $\t$'s, and that we will have to solve our constraints $D^2I=\ol{D}^2I=0$ within each sector of fixed charges.  Note also that we can only write down candidate structures when the total R-charge is even, and that a reality condition will relate terms with negative total R-charge to those with positive total R-charge.  This then splits the computation into several sectors.  We will briefly go through them.

In the sector with R-charges $(R,R')=(2,2)$, the only possible structure has the form $I_{(2,2)}=A(p^2)\,\t^2\,{\t'}^2$ for some function $A(p^2)$, but it is easy to check that the condition $D^2I_{(2,2)}=0$ is satisfied only for $A(p^2)=0$.

Similarly, in the sectors $(2,-2)+(-2,2)$ (with structure $(\t^2{\ol{\t}'}^2+\ol{\t}^2{\t'}^2)$) and $(2,0)+(0,2)$ (with structures $(\t^2+{\t'}^2)$, $p_\m(\t^2\t'\s^\m\ol{\t}'-{\t'}^2\t\s^\m\ol{\t})$, and $(\t^2{\t'}^2{\ol{\t}'}^2+\t^2\ol{\t}^2{\t'}^2)$) there are no nontrivial solutions to the condition $D^2I=0$.

In the sector $(1,1)$, there are three structures that one can write down ($\t\t'$, $p_\m(\t^2\t'\s^\m\ol{\t}-{\t'}^2\t\s^\m\ol{\t}')$, and $\t^2{\t'}^2\ol{\t}\ol{\t}'$), and precisely one linear combination of them satisfies $D^2I=0$, namely
\begin{equation}
I_{(1,1)}=F_1(p^2)\lp\t\t'+\hlf p_\m\,\lp\t^2\,\t'\s^\m\ol{\t}-{\t'}^2\,\t\s^\m\ol{\t}'\rp-\frac{1}{4}p^2\,\t^2\,{\t'}^2\,\ol{\t}\ol{\t}'\rp.
\end{equation}
Here $F_1(p^2)$ is an arbitrary complex-valued function, and to obtain a real function $I$ we will need to also add the complex conjugate $I_{(-1,-1)}=\ol{I_{(1,1)}}$.  Note that $\ol{D}^2I_{(1,1)}=0$ automatically.

In the sector $(1,-1)+(-1,1)$ there are initially four structures, but only one combination survives the constraint $D^2I=\ol{D}^2I=0$,
\begin{equation}
\begin{split}
I_{(1,-1)+(-1,1)} &= F_2(p^2)\lp p_\m\,\lp\t\s^\m\ol{\t}'-\t'\s^\m\ol{\t}\rp+\hlf p^2\,\lp\t^2\,\ol{\t}\ol{\t}'+{\t'}^2\,\ol{\t}\ol{\t}'+\t\t'\,{\ol{\t}'}^2+\t\t'\,\ol{\t}^2\rp\qquad\right.\\
& \qquad\left.+\frac{1}{4}p^2p_\m\,\lp\t^2\,{\ol{\t}'}^2\,\t'\s^\m\ol{\t}-\ol{\t}^2\,{\t'}^2\,\t\s^\m\ol{\t}'\rp\rp,
\end{split}
\end{equation}
where $F_2(p^2)$ is an arbitrary real function.

Finally in the sector $(0,0)$ there are initially seven structures one can write down, but imposing $D^2I=\ol{D}^2I=0$ reduces this down to two independent contributions,
\begin{equation}
\begin{split}
I_{(0,0)} =& F_3(p^2)\lp 1+\frac{1}{4}p^2\,\lp\t^2\ol{\t}^2+{\t'}^2{\ol{\t}'}^2\rp+\frac{1}{16}p^4\,\t^2\,\ol{\t}^2\,{\t'}^2\,{\ol{\t}'}^2\rp\\
&-F_4(p^2)\lp p^2\eta_{\m\n}-p_\m p_\n\rp\,\t\s^\m\ol{\t}\,\t'\s^\n\ol{\t}'.
\end{split}
\end{equation}
$F_3(p^2)$ and $F_4(p^2)$ will both be real functions of the square of the momentum.

The most general solution is then the combination
\begin{equation}
I(p,\t,\ol{\t},\t',\ol{\t}')=\lp I_{(1,1)}+I_{(-1,-1)}\rp+I_{(1,-1)+(-1,1)}+I_{(0,0)},
\end{equation}
which depends on the the real-valued functions $F_2$, $F_3$, and $F_4$, and the complex-valued function $F_1$.  Note that since the supercurrent $\mathcal{J}$ is dimensionless, the functions $F_2$, $F_3$, and $F_4$ should also be dimensionless, while the function $F_1$ should have dimensions of mass.

The supersymmetric result \eqref{susyresult} is recovered, by setting $F_{1}=0$ and $F_{3}=F_{4}= -F_{2} = C_{\text{SUSY}}$.

\subsection{Comparison with Meade et al.}

We can expand the supercurrent in components as
\begin{equation}
\mathcal{J}(p,\t,\ol{\t})=J(p)+i\t j(p)-i\ol{\t}\ol{\jmath}-j_\m\,\t\s^\m\ol{\t}+\frac{i}{2}p_\m\,\t^2\, j\s^\m\ol{\t}+\frac{i}{2}p_\m\,\ol{\t}^2\,\t\s^\m\ol{\jmath}+\frac{1}{4}p^2\,\t^2\,\ol{\t}^2\, J.
\end{equation}
Note that the conditions $D^2\mathcal{J}=\ol{D}^2\mathcal{J}=0$ are then satisfied provided that $p^\m j_\m=0$.

Expanding the two-point function in components, we find (dropping the overall factor of $(2\pi)^4\d^4(p+p')$)
\begin{equation}
\begin{split}
\left\langle\mathcal{J}(p)\,\mathcal{J}'(-p)\right\rangle &= \left\langle J(p)\, J(-p)\right\rangle\ls 1+\frac{1}{4}p^2\,\lp\t^2\,\ol{\t}^2+{\t'}^2\,{\ol{\t}'}^2\rp+\frac{1}{16}p^4\,\t^2\,\ol{\t}^2\,{\t'}^2\,{\ol{\t}'}^2\rs\\
& +\left\langle j_\al(p)\, j_\beta(-p)\right\rangle\ls\t^\al{\t'}^\beta-\frac{1}{2}p_\m\,\t^\al\,{\t'}^2\,\e^{\beta\g}\s^\m_{\g\dot\g}{\ol{\t}'}^{\dot\g}+\frac{1}{2}p_\m\,\t^2\,\e^{\al\g}\s^\m_{\g\dot\g}\ol{\t}^{\dot\g}\,{\ol{\t}'}^{\dot\beta}\right.\\
& \qquad\left.-\frac{1}{4}p_\m p_\n\,\t^2\,{\t'}^2\,\e^{\al\g}\s^\m_{\g\dot\g}\ol{\t}^{\dot\g}\,\e^{\beta\d}\s^\n_{\d\dot\d}{\ol{\t}'}^{\dot\d}\rs\\
& +\left\langle j_\al(p)\,\ol{\jmath}_{\dot\beta}(-p)\right\rangle\ls\t^\al{\ol{\t}'}^{\dot\beta}+\frac{1}{2}p_\m\,\t^\al\,{\ol{\t}'}^2\,{\t'}^\g\,\e^{\dot\beta\dot\g}\s^\m_{\g\dot\g}+\frac{1}{2}p_\m\,\t^2\,\e^{\al\g}\s^\m_{\g\dot\g}\ol{\t}^{\dot\g}\,{\ol{\t}'}^{\dot\beta}\right.\\
& \qquad\left.+\frac{1}{4}p_\m p_\n\,\t^2\,{\ol{\t}'}^2\,\e^{\al\g}\s^\m_{\g\dot\g}\ol{\t}^{\dot\g}\,{\t'}^\d\,\e^{\dot\beta\dot\d}\s^\n_{\d\dot\d}\rs\\
& +\left\langle j_\m(p)\, j_\n(-p)\right\rangle\ls\t\s^\m\ol{\t}\,\t'\s^\n\ol{\t}'\rs\label{mcJmcJ}\\
& +\left\langle\ol{\jmath}_{\dot\beta}(p)\,j_\al(-p)\right\rangle\ls\ol{\t}^{\dot\beta}{\t'}^\al-\hlf p_\m\,\ol{\t}^{\dot\beta}\,{\t'}^2\,\e^{\al\g}\s^\m_{\g\dot\g}{\ol{\t}'}^{\dot\g}-\hlf p_\m\,\ol{\t}^2\,\t^\g\e^{\dot\beta\dot\g}\s^\m_{\g\dot\g}\,{\t'}^\al\right.\\
& \qquad\left.+\frac{1}{4}p_\m p_\n\,\ol{\t}^2\,{\t'}^2\,\t^\g\e^{\dot\beta\dot\g}\s^\m_{\g\dot\g}\,\e^{\al\d}\s^\n_{\d\dot\d}{\ol{\t}'}^{\dot\d}\rs\\
& +\left\langle\ol{\jmath}_{\dot\al}(p)\,\ol{\jmath}_{\dot\beta}(-p)\right\rangle\ls\cdots\rs.
\end{split}
\end{equation}

The authors of \cite{Meade:2008wd} then make the definitions
\begin{equation}\label{jj}
\begin{split}
\left\langle J(p)\, J(-p)\right\rangle = & \wt{C}_0(p^2),\\
\left\langle j_\al(p)\,\ol{\jmath}_{\dot\beta}(-p)\right\rangle = & -p_\m\s^\m_{\al\dot\beta}\,\wt{C}_{1/2}(p^2),\\
\left\langle j_\m(p)\, j_\n(-p)\right\rangle = & -\lp p^2\eta_{\m\n}-p_\m p_\n\rp\wt{C}_1(p^2),\\
\left\langle j_\al(p)\, j_\beta(-p)\right\rangle = & \e_{\al\beta}\,m\wt{B}_{1/2}(p^2).\end{split}
\end{equation}
Here $m$ is some mass scale for the messenger sector.  In subsequent sections we will take it to be the lowest component VEV of a spurion superfield $M$, but for now we can consider the combination $m\wt{B}_{1/2}(p^2)$ to be an arbitrary function of mass dimension one. 

Comparing (\ref{mcJmcJ}), (\ref{jj}), and the results from the previous section, we find complete agreement, provided we identify
\begin{equation}
F_1=m\wt{B}_{1/2},\qquad F_2=-\wt{C}_{1/2},\qquad F_3=\wt{C}_0,\qquad F_4=\wt{C}_1.
\end{equation}

We will adopt the notation of \cite{Meade:2008wd} for the remainder of the paper.

In what follows, we will need to avail ourselves of the fact that supersymmetry is dynamically, rather than explicitly, broken in the hidden sector. This determines the UV behaviour of the functions $\wt{C}_{s}$. In particular, we will need to make use of some facts about the asymptotics of the functions $\wt{C}_{s}$. Since supersymmetry is restored at short distances,
\begin{equation}
\wt{C}_{s}(p^2)=C_{\text{SUSY}}(p^2)+f_{s}\left(\frac{p^2}{|m|^2}\right),
\end{equation}
where $\wt{B}_{1/2}(y)$ and $f_{s}(y)\to 0$ as $y\to\infty$ and are analytic near $y=0$.

\section{General Results}
\label{GeneralResults}

\subsection{Connection between the current two-point functions and the effective action}
\label{GeneralConnection}

The gauge current superfield will act as a source for the Standard Model vector superfields,
\begin{equation}
\mathcal{L}_{int}=2\int d^4\t\mathcal{J}V+\cdots=JD-j\la-\ol{\jmath}\ol{\la}-j^\m v_\m+\cdots,
\end{equation}
where $\cdots$ refers to terms that are higher than linear order in the vector multiplet.  Note that the normalization here differs from \cite{Meade:2008wd} by a factor of the gauge coupling $g$.  Their expansion, which is in powers of the standard model gauge couplings, will correspond, here, to an expansion in the number of gauge multiplet fields.  If there are several gauge factors, such as the three gauge groups of the MSSM, then there will be a supercurrent and a vector superfield for each one, and we will label the gauge groups by subscripts, $\mathcal{J}_r$, $V_r$, etc.  For the case of the MSSM, we will let $r=1,2,3$ refer to $U(1)$, $SU(2)$, $SU(3)$, respectively.  Also we will use the $SU(5)$ normalization for the $U(1)$ coupling $g_1$ (which differs by a factor of $\sqrt{3/5}$ from the hypercharge normalization).

Upon integrating out the messenger sector and picking out the terms quadratic in the gauge couplings and quadratic in the MSSM gauge fields, we find that the only effect of the messenger sector is through the two-point functions discussed above, which appear in the effective action\footnote{This is for a single abelian vector multiplet.  The generalization to nonabelian gauge groups is straightforward \cite{Meade:2008wd}.} as
\begin{equation}
\begin{split}
\label{TwoPointInLeff}
\d\mathcal{L}_{eff} =& \hlf\wt{C}_0(p^2)D^2-\wt{C}_{1/2}(p^2)i\la\s^\m\p_\m\ol{\la}+\hlf\wt{C}_1(p^2) v^\m\lp\eta_{\m\n}\square-\p_\m\p_\n\rp v^\n\\
& \qquad-\hlf\lp m\wt{B}_{1/2}(p^2)\la\la+\mathrm{c.c}\rp.
\end{split}
\end{equation}
This is in addition to the stanard kinetic terms
\be
\mathcal{L}_{kin}=\frac{1}{4g^2}\lp\int d^2\t\, W^\al W_\al+\mathrm{c.c}\rp.
\ee
From this we see that the gauginos receive a mass at tree level,
\begin{equation}
m_\lambda^{(r)}=g_r^2m\wt{B}_{1/2}^{(r)}(0),
\end{equation}
where we now allow for different couplings and two-point functions for different gauge groups. The functions $\wt{B}_{1/2}^{(r)}(p^2)$ are a-priori arbitrary. Thus there is no hint of gaugino mass unification in this framework, and no reason to expect the bino to be the lightest of the gauginos.

The sfermions receive a mass at one-loop, given by
\begin{equation}\label{SfermionBad}
m_{\wt{f}}^2=\sum_{r=1}^3g_r^4c_2(f;r)A_r,
\end{equation}
where
\begin{equation}\label{ArBad}
\begin{split}
A_r =& -\int\frac{d^4p}{(2\pi)^4}\frac{1}{p^2}\lp 3\wt{C}^{(r)}_{1}(p^2)-4\wt{C}^{(r)}_{1/2}(p^2)+\wt{C}^{(r)}_{0}(p^2)\rp\\
=& -\frac{|m|^2}{16\pi^2}\int_0^\infty dy\lp 3\wt{C}^{(r)}_{1}(y)-4\wt{C}^{(r)}_{1/2}(y)+\wt{C}^{(r)}_{0}(y)\rp.
\end{split}
\end{equation}
Here we have made the substitution $y=p^2/|m|^2$, where, again, $m$ is the mass scale characterizing the hidden sector.  $c_2(f;r)$ is the quadratic Casimir for the representation of $f$ under the gauge group $r$ (equal to $(N^2-1)/(2N)$ for the fundamental representation of $SU(N)$).  We have also assumed that $\mathcal{J}$ had no non-vanishing one-point function, as this would also contribute to the sfermion masses.  Note also that we are working to lowest order in the MSSM gauge couplings.  For instance there is also a contributing diagram with one-loop of gauginos and two insertions of $\wt{B}_{1/2}(p^2)$, but we ignore it since it is order $g_r^6$.

In this framework, there's no particular reason to expect the $A_r$ to be positive. Indeed, as noted in the introduction, simple variants of anomaly mediation give examples where the slepton masses turn out to be tachyonic. Moreover, for the integral in \eqref{ArBad} to be UV convergent (as we expect) would require stronger assumptions about large-$y$ behaviour of the $C_s(y)$ than we have hitherto imposed. We will see however that in the case of SUSY breaking by a single spurion field that both these worries evaporate. In \S\ref{multispurion}, we will comment briefly on the generalization to multiple spurions.

The expressions above for gaugino and sfermion masses are valid at the scale $m$ after integrating out the messenger fields.  To obtain the masses at lower scales, one runs these down, using the usual renormalization group techniques~\cite{Giudice:1997ni,Cohen:2006qc}.  

\subsection{F-term breaking by a single spurion field}
\label{SingleSpurionResults}

We will now focus our attention on a more specific situation (though still applicable to a large class of models).  We assume that supersymmetry is broken only by the auxiliary VEV of a single spurion chiral superfield, $M=m+\t^2F$.  We will consider the most general effective action involving the superfields $M$, $\ol{M}$, $W^\al$, $\ol{W}^{\dot\al}$, and which is quadratic in the gauge fields.

The first simplifying assumption that will allow us to make progress is that    
there is a small parameter, $|F/m^2|\ll 1$. We will thus expand our             
effective action in powers of this small parameter, keeping terms of order      
$F$, $\overline{F}$ and $|F|^2$, but dropping terms of order $F|F|^2$,          
$\overline{F}|F|^2$ and higher.                                                 
                                                                                
Our second simplifying assumption is that the dynamics of the hidden sector     
has a chiral $U(1)$ symmetry, under which $M$ and $\overline{M}$ rotate with    
opposite phases. In concrete realizations, the existence of such a $U(1)$ symmetry is fairly commonplace. If $S$ is weakly-coupled, with messenger chiral multiplets, $M$ is the messenger mass, and the $U(1)$ is the usual chiral rotations acting on the fermions. In a strongly-coupled theory, with a dynamically-broken chiral symmetry, $M$ is $\Lambda_{\text{QCD}}$, the dynamical scale of that theory.

We assume that this symmetry is unbroken, except for the       
anomaly. That is, all terms in the effective action are $U(1)$-invariant,       
\emph{except} for                                                               
$$                                                                              
  \int d^4 x \left(\int d^2\theta \log(M) Tr WW + \text{h. c.}\right)           
$$                                                                              
Invariance of this term requires that a phase rotation on $M$ be accompanied    
by a shift in the $\Theta$-angle. More precisely, including the kinetic term     
for the gauge fields,                                                           
\begin{equation}\label{anom}                                                    
S_{\text{eff}} = \int d^4 x \left(\int d^2\theta\left(\tfrac{1}{4g^2(\mu)}-\tfrac{i\Theta}{32\pi^2}     
-\tfrac{c}{2} \log(M/\mu) \right) Tr WW + \text{h. c.} \right) +U(1)\text{-invariant}       
\end{equation}                                                                  
and a phase rotation on $M$ is accompanied by a shift of $\Theta$.
                                                                       
The first thing to notice is that the anomaly coefficient is \emph{the same}    
as the jump in the Standard Model $\beta$-function coefficient from             
integrating out the messenger sector. The former involves the coefficient of    
$F\wedge F$ in \eqref{anom}; the latter involves the coefficient of             
$F_{\mu\nu}F^{\mu\nu}$. Holomorphy in $M$ relates these two.                    
                                                                                
Moreover, we will see that the gluino mass,                                                      
\begin{equation}\label{GauginoMass}                                                                
 m_\lambda^{(r)} = g_r^2 m \tilde{B}_{1/2}^{(r)}(0) = g_r^2 c^{(r)}  \tfrac{F}{m} +           
O(g_r^2\tfrac{F}{m}\tfrac{|F|^2}{|m|^4})                                                                       
\end{equation}                                                                  
is given, to leading order (and \emph{only} to leading order) by this term.     
                                                                                
If the gauge mediation model has coupling constant unification (an extra        
assumption), then as discussed above, all of the $c^{(r)}$ are equal, and the ratios of gluino          
masses are equal, at this order, to the ratios of gauge-couplings squared       
(in the $SU(5)$ normalization). This is known as ``gluino mass unification",     
and we see that it emerges, in the limit of weak               
supersymmetry breaking ($|F/m^2|\ll 1$), quite generally, in this                
model-independent context.                                                      

In fact, we will show in the next section that these assumptions give us much stronger relations.  In this context we can expand each of the functions $\wt{C}_s$ (when there's no need for it, we will drop the gauge group index, $r$) as
\begin{equation}
\wt{C}_s(p^2)=C_{\mathrm{SUSY}}(p^2)+\sum_{k=1}^\infty\frac{|F|^{2k}}{|m|^{4k}}\mathcal{C}_s^{(2k)}(y),
\end{equation}
where as above,
\begin{equation}
C_{\mathrm{SUSY}}(p^2)=c\ln\frac{\La^2}{|m|^2}+f(p^2/|m|^2),
\end{equation}
for some cutoff-independent function, $f(y)$, analytic near $y=0$.

Similarly, we can write (R-charge considerations force us to have one more $F$ than $\ol{F}$)
\begin{equation}\label{Bexpansion}
\wt{B}_{1/2}(p^2)=\frac{F}{m^2}\sum_{k=0}^\infty\frac{|F|^{2k}}{|m|^{4k}}\mathcal{B}^{(2k+1)}(y).
\end{equation}
The functions $\mathcal{C}_s^{(2k)}(y)$ and $\mathcal{B}^{(2k+1)}(y)$ should be analytic near $y=0$,  and finite as $y\to\infty$.

With this set of assumptions and definitions, we shall then show below that
\begin{equation}\label{B1result}
\mathcal{B}^{(1)}(p^2)=-|m|^2\frac{\p}{\p(|m|^2)}C_{\mathrm{SUSY}}(p^2).
\end{equation}
We already saw above in \eqref{GauginoMass} that the zero momentum part of this relation was satisfied, and that was sufficient to determine the gaugino masses.  The full momentum dependence enters into the computation of the scalar A-terms.

We will also show that
\begin{equation}
\begin{split}
3\mathcal{C}_1^{(2)}(p^2)-4\mathcal{C}_{1/2}^{(2)}(p^2)+\mathcal{C}_0^{(2)}(p^2) =& 2|m|^2\frac{\p}{\p(p^2)}\lp p^2\frac{\p}{\p(p^2)}C_{\mathrm{SUSY}}(p^2)\rp\\
=& 2\frac{\p}{\p y}\lp y\frac{\p}{\p y}C_{\mathrm{SUSY}}(y)\rp.
\end{split}
\end{equation}
It then follows, using the results above, that
\begin{equation}\label{ArResult}
\begin{split}
A_r =& -\frac{|m|^2}{16\pi^2}\int_0^\infty dy\lp 3\wt{C}^{(r)}_{1}(y)-4\wt{C}^{(r)}_{1/2}(y)+\wt{C}^{(r)}_{0}(y)\rp\\
=&-\frac{|F|^2}{8\pi^2|m|^2}\lp y\left.\frac{\p}{\p y}C_{\mathrm{SUSY}}\right|_0^\infty+\mathcal{O}(\frac{|F|^4}{|m|^6}\rp\\
=& \frac{c^{(r)}}{8\pi^2}\frac{|F|^2}{|m|^2}+\mathcal{O}\left(\frac{|F|^4}{|m|^6}\right).
\end{split}
\end{equation}
To this order, the $A_r$ are independent of any of the details of the momentum dependence of the structure functions and are determined entirely by the shifts in the $\beta$-function, $c^{(r)}$.  In particular, 
the $A_r$, and hence the sfermion masses \eqref{SfermionBad}, are manifestly positive.

\section{Higher Derivative Terms}
\label{HigherDerivativeTerms}

We would like to write down all gauge invariant terms which can be constructed out of the field strengths $W^\al=-\frac{1}{4}\ol{D}^2D^\al V$ and $\ol{W}^{\dot\alpha}=-\frac{1}{4}D^2\ol{D}^{\dot\al}V$ (and is quadratic in these quantities), a spurion superfield $M=m+\t^2F$, and derivatives thereof.  We want our terms to have mass dimension two and R-charge zero so that we can integrate over $d^4\t$ and add it to the effective action (we assume that $M$ has zero R-charge, and hence that the R-symmetry is spontaneously broken by the VEV of $F$ \cite{Nelson:1993nf}).  Finally, we also assume a global $U(1)$ symmetry under which only $M$ is charged, which forces us to take equal numbers of $M$ and $\ol{M}$.  Note that we could use this global symmetry to fix $m$ to be real, but we won't yet make that choice.  We can then work at a fixed order in $|M|^{-1}$.  

\subsection{Leading order}

We will now consider higher derivative terms of order $|M|^{-2}$.  We can have terms with two $W$ and two $D$, which can either hit a $W$ or an $M$, or we can have terms with one $W$, one $\ol{W}$, and either one $D$ (hitting $W$ or $M$) and one $\ol{D}$ (hitting $\ol{W}$ or $\ol{M}$), or a partial derivative $\p$ (hitting either $W$ or $\ol{W}$).  There can also be terms with two $\ol{W}$ and two $\ol{D}$, but these will be hermitian conjugates of the first class of terms.  In appendix \ref{LOTerms} we present a full list of possible terms and the equivalences between them in order to find the four independent real terms which can be added to the effective action, but we can present a brief derivation here.

We note that there are various identities which hold and which allow us to identify some terms with others.  For example we have an identity
\begin{equation}
\label{EpsilonIdentity}
\e^{\al\beta}\e^{\g\d}+\e^{\al\d}\e^{\beta\g}=\e^{\al\g}\e^{\beta\d},
\end{equation}
and similarly for dotted indices.

We also have
\begin{equation}
\label{DWIdentity}
D^\al W_\al=\ol{D}_{\dot\al}\ol{W}^{\dot\al}.
\end{equation}
Differentiating this relation gives us
\begin{equation}
\label{D^2WIdentity}
D^2W_\al=4i\s^\m_{\al\dot\al}\p_\m\ol{W}^{\dot\al},\qquad\ol{D}^2\ol{W}_{\dot\al}=-4i\s^\m_{\al\dot\al}\p_\m W^\al.
\end{equation}

Consider now the terms with two $W$ and two $D$.  Using (\ref{D^2WIdentity}) we may assume that both $D$'s do not hit the same $W$.  Moreover, if each $W$ is differentiated, then we can integrate by parts (recall that our terms will be integrated over $d^4x d^4\t$) to relate such a term to one in which one $D$ hits an $M$.  Equation (\ref{EpsilonIdentity}) shows that there are two independent terms in this class, which we can take to be
\begin{equation}
|M|^{-2}\lp M^{-1}D^\al M\rp\lp W_\al D^\beta W_\beta\rp,\qquad|M|^{-2}\lp M^{-1}D^\al M\rp\lp W^\beta D_\al W_\beta\rp.
\end{equation}
Now suppose neither $D$ acts on a $W$.  If they both hit the same $M$, then we could integrate by parts to relate it to terms discussed above (where one $D$ hits a $W$) or a term in which each $D$ hits a different $M$,
\begin{equation}
|M|^{-2}\lp M^{-1}D^\al M\rp\lp M^{-1}D_\al M\rp\lp W^\beta W_\beta\rp.
\end{equation}

Finally, we turn to the terms with one $W$ and one $\ol{W}$.  If we have a $D$ and a $\ol{D}$, then by integrating the $\ol{D}$ by parts if necessary, we will have the structure $\ol{D}_{\dot\al}\ol{W}^{\dot\al}$ appearing, and by (\ref{DWIdentity}) we are back to the terms just discussed.  The only remaining possibility is to have a partial derivative.  By integrating it by parts we may assume that it acts on $\ol{W}$, and so the only remaining independent term is
\begin{equation}
i|M|^{-2}\s^\m_{\al\dot\al}W^\al\p_\m\ol{W}^{\dot\al}.
\end{equation}

Altogether then, the most general contribution to the effective action at this order is
\begin{equation}
\begin{split}
\d\mathcal{L}_{eff} =& \int d^4\t|M|^{-2}\ls c_1\lp M^{-1}D^\al M\,W_\al D^\beta W_\beta+\mathrm{h.c.}\rp+c_2\lp M^{-1}D^\al M\,W^\beta D_\al W_\beta+\mathrm{h.c.}\rp\right.\\
& \qquad\left.+c_3\lp M^{-2} D^\al MD_\al M\,W^\beta W_\beta+\mathrm{h.c.}\rp+c_4\lp i\s^\m_{\al\dot\al}W^\al\p_\m\ol{W}^{\dot\al}+\mathrm{h.c.}\rp\rs.
\end{split}
\end{equation}
Expanding this into component fields and doing the $d^4\t$ integration, we find
\begin{equation}
\begin{split}
\d S_{eff} =& \int d^4x\ls 4c_1\frac{|F|^2}{|m|^4}\lp 2D^2-i\la\s^\m\p_\m\ol{\la}\rp\right.\\
& \qquad\left.+4c_2\frac{|F|^2}{|m|^4}\lp D^2-2i\la\s^\m\p_\m\ol{\la}+v^\m\lp\eta_{\m\n}\square-\p_\m\p_\n\rp v^\n\rp\right.\\
& \qquad\left.+4c_3\frac{|F|^2}{|m|^4}\lp\frac{F}{m}\la\la+\frac{\ol{F}}{\ol{m}}\ol{\la}\ol{\la}\rp\right.\\
& \qquad\left.+2c_1\lp-D\frac{\square}{|m|^2}D+2i\la\s^\m\frac{\square}{|m|^2}\p_\m\ol{\la}-v^\m\lp\eta_{\m\n}\square^2-\square\p_\m\p_\n\rp v^\n\right.\right.\\
& \qquad\left.\qquad\left.-\frac{F}{m}\la\frac{\square}{|m|^2}\la-\frac{\ol{F}}{\ol{m}}\ol{\la}\frac{\square}{|m|^2}\ol{\la}+\frac{|F|^2}{|m|^4}i\la\s^\m\p_\m\ol{\la}\rp\rs
\end{split}
\end{equation}

Now comparing with (\ref{TwoPointInLeff}), we can conclude that the terms we have listed contribute as
\begin{equation}
\begin{split}
\d\wt{C}_0 =& 4c_4\frac{p^2}{|m|^2}+8\lp 2c_1+c_2\rp\frac{|F|^2}{|m|^4},\\
\d\wt{C}_{1/2} =& 4c_4\frac{p^2}{|m|^2}+2\lp-c_4+2c_1+4c_2\rp\frac{|F|^2}{|m|^4},\\
\d\wt{C}_1 =& 4c_4\frac{p^2}{|m|^2}+8c_2\frac{|F|^2}{|m|^4},\\
\d\wt{B}_{1/2} =& -4c_4\frac{F}{m^2}\frac{p^2}{|m|^2}-8c_3\frac{F}{m^2}\frac{|F|^2}{|m|^4}.
\end{split}
\end{equation}

We note that at order $F^0$ only $c_4$ contributes, and it preserves the supersymmetry condition $\wt{C}_0^{(0)}=\wt{C}_{1/2}^{(0)}=\wt{C}_1^{(0)}=C_{\mathrm{SUSY}}$.  We also note that the contribution to $\mathcal{B}^{(1)}$ depends only on $c_4$, that is it is determined by the contribution to $C_{\mathrm{SUSY}}$.  Similarly, the combination that appears in the computation of sfermion masses is given by
\begin{equation}
\d\lp 3\wt{C}_1-4\wt{C}_{1/2}+\wt{C}_0\rp=8c_4\frac{|F|^2}{|m|^4},
\end{equation}
and it too depends only on the supersymmetric contribution.  On the other hand, $c_3$ contributes only to the zero momentum piece of $\mathcal{B}^{(3)}$, so the latter is completely independent.  Similarly, the other two linearly independent combinations of the $\mathcal{C}_s^{(2)}$ can be set to anything by adjusting $c_1$ and $c_2$.

Next we will see how these features persist to all orders in $|M|^{-1}$.

\subsection{All-order calculations}

Let us now consider terms of order $|M|^{-2n}$ for $n\ge 2$, but where we will restrict our attention to those that contribute to $\wt{B}_{1/2}$ at order $F$ or to $\wt{C}_s$ at order $1$ or order $|F|^2$.

First let us consider the terms with two $W^\al$.  Such a term will be built out of $k$ partial derivatives $\p$, for $0\le k\le 2n-2$, $2n-k$ covariant derivatives $D$, and $2n-2-k$ derivatives $\ol{D}$.  Here the partial derivatives act only on the $W$'s, $\ol{D}$ act only on $\ol{M}$, and $D$ can act on either $W$ or $M$.  Using (\ref{D^2WIdentity}), we can assume that each $W$ is hit by at most one $D$.  By integration by parts we can further assume that all the $\p$ act on the same $W$, and that the other $W$ is not hit by any $D$.  This leaves $2n-k-1$ $D$'s which act on $M$'s, but in order to get contributions with at most one $F$, we can only act on a single $M$, hence we need $2n-k-1\le 2$.  These considerations lead us to the following possible independent terms (where we also use (\ref{EpsilonIdentity}) to drop terms):
\begin{equation}
\begin{split}
K_1 =& i\left|M\right|^{-2n-2}\s^\m_{\al\dot\al} \lp D^2M\rp\lp\ol{D}^{\dot\al}\ol{M}\rp\lp W^\al \square^{n-2}\p_\m D^\beta W_\beta\rp+h.c.,\\
K_2 =& i\left|M\right|^{-2n-2}\s^\m_{\al\dot\al} \lp D^2M\rp\lp\ol{D}^{\dot\al}\ol{M}\rp\lp W^\beta \square^{n-2}\p_\m D^\al W_\beta\rp+h.c.,\\
K_3 =& \left|M\right|^{-2n-2}\lp\ol{M} D^\al M\rp\lp W_\al\square^{n-1} D^\beta W_\beta\rp+h.c.,\\
K_4 =& \left|M\right|^{-2n-2}\lp\ol{M} D^\al M\rp\lp W^\beta\square^{n-1} D_\al W_\beta\rp+h.c.,\\
K_5 =& \left|M\right|^{-2n-2}\lp\ol{M} D^2M\rp\lp W^\al\square^{n-1} W_\al\rp+h.c.
\end{split}
\end{equation}
Here, of course, $\square=\p^\m\p_\m$.

Turning next to terms with one $W$ and one $\ol{W}$, we can have again $k$ partial derivatives $\p$,with $0\le k\le 2n-1$, along with $2n-k-1$ each of $D$ and $\ol{D}$.  By integrations by parts we can arrange for all the $\p$ to hit $\ol{W}$, and for all $D$ to hit $M$ and all $\ol{D}$ to hit $\ol{M}$.  This then leads to
\begin{equation}
\begin{split}
K_6 =& i\left|M\right|^{-2n-2}\s^\m_{\al\dot\al}\lp D^2M\rp\lp\ol{D}^2\ol{M}\rp\lp W^\al\square^{n-2}\p_\m\ol{W}^{\dot\al}\rp+h.c.,\\
K_7 =& \left|M\right|^{-2n-2}\s^\m_{\al\dot\al}\s^\n_{\beta\dot\beta}\lp D^\al M\rp\lp\ol{D}^{\dot\al}\ol{M}\rp\lp W^\beta\square^{n-2}\p_\m\p_\n\ol{W}^{\dot\beta}\rp+h.c.,\\
K_8 =& \left|M\right|^{-2n-2}\s^\m_{\al\dot\al}\s^\n_{\beta\dot\beta}\lp D^\al M\rp\lp\ol{D}^{\dot\beta}\ol{M}\rp\lp W^\beta\square^{n-2}\p_\m\p_\n\ol{W}^{\dot\al}\rp+h.c.,\\
K_9 =& i\left|M\right|^{-2n}\s^\m_{\al\dot\al}\lp W^\al\square^{n-1}\p_\m\ol{W}^{\dot\al}\rp+h.c..
\end{split}
\end{equation}

Then proceeding as above, we perform the $d^4\t$ integration on $\sum c_i'K_i$ and find the contributions
\begin{equation}
\begin{split}
\wt{C}_0 =& -4\frac{p^{2n}}{|m|^{2n}}\lp-1\rp^nc_9'+8\frac{|F|^2p^{2n-2}}{|m|^{2n+2}}\lp-1\rp^n\lp-4c_1'-8c_2'-2nc_3'-nc_4'-2nc_5'\right.\\
& \qquad\left.+8c_6'+c_7'-c_8'\rp,\\
\wt{C}_{1/2} =& -4\frac{p^{2n}}{|m|^{2n}}\lp-1\rp^nc_9'+2\frac{|F|^2p^{2n-2}}{|m|^{2n+2}}\lp-1\rp^n\lp-16c_1'-32c_2'-2nc_3'-4nc_4'-8nc_5'\right.\\
& \qquad\left.+32c_6'+4c_7'+2c_8'+n^2c_9'\rp-32\lp n+1\rp^2\frac{|F|^4p^{2n-4}}{|m|^{2n+4}}\lp-1\rp^nc_6',\\
\wt{C}_1 =& -4\frac{p^{2n}}{|m|^{2n}}\lp-1\rp^nc_9'+8\frac{|F|^2p^{2n-2}}{|m|^{2n+2}}\lp-1\rp^n\lp-4c_1'-8c_2'-nc_4'-2nc_5'+8c_6'+c_7'+c_8'\rp,\\
\wt{B}_{1/2} =& -4n\frac{F}{m^2}\frac{p^{2n}}{|m|^{2n}}\lp-1\rp^nc_9'+16\lp n+1\rp\frac{F}{m^2}\frac{|F|^2p^{2n-2}}{|m|^{2n+2}}\lp-1\rp^n\lp 2c_1'+4c_2'+nc_5'-4c_6'\rp.
\end{split}
\end{equation}

We see once again that the SUSY contribution is uniquely determined by $c_9'$, and this also uniquely fixes the order $F$ contribution to $B_{1/2}$.  Similarly, the combination relevant for sfermion masses is
\begin{equation}
3\wt{C}_1-4\wt{C}_{1/2}+\wt{C}_0=-8n^2\frac{|F|^2p^{2n-2}}{|m|^{2n+2}}\lp-1\rp^nc_9'+\mathcal{O}(\frac{|F|^4p^{2n-4}}{|m|^{2n+4}}),
\end{equation}
so the order $|F|^2$ piece also depends only on $c_9'$.

In fact, we can write these relations to all orders more succinctly.  If we write
\begin{equation}
C_{\mathrm{SUSY}}(p^2)=\sum_{k=0}^\infty\g_k\frac{p^{2k}}{|m|^{2k}},
\end{equation}
where for $k\ge 1$, $\g_k$ is just a number, while $\g_0$ can be logarithmically divergent, that is it can be an affine linear function of $\ln(\Lambda^2/|m|^2)$.  Then to order $F$ we find,
\begin{equation}
\begin{split}
m\wt{B}_{1/2}(p^2) =& m\wt{B}_{1/2}(0)+\frac{F}{m}\sum_{k=1}^\infty k\g_k\frac{p^{2k}}{|m|^{2k}}+\mathcal{O}\left(\frac{F}{m}\frac{|F|^2}{|m|^4}\right)\\
=& m \wt{B}_{1/2}(0)+\frac{F}{m}p^2\frac{\p}{\p(p^2)}C_{\mathrm{SUSY}}(p^2)+\mathcal{O}\left(\frac{F}{m}\frac{|F|^2}{|m|^4}\right)\\
=& -\frac{F}{m}|m|^2\frac{\p}{\p(|m|^2)}C_{\mathrm{SUSY}}+\mathcal{O}\left(\frac{F}{m}\frac{|F|^2}{|m|^4}\right).\label{BRelation}
\end{split}
\end{equation}
In the final step we have also used the relationship between $\wt{B}_{1/2}(0)$ and the coefficient of $\ln(\Lambda^2/|m|^2)$ in $C_{\mathrm{SUSY}}$.

Similarly, we can write, at order $|F|^2$,
\begin{equation}
\label{CRelation}
3\wt{C}_1(p^2)-4\wt{C}_{1/2}(p^2)+\wt{C}_0(p^2)=\frac{2|F|^2}{|m|^2}\frac{\p}{\p(p^2)}\lp p^2\frac{\p}{\p(p^2)}C_{\mathrm{SUSY}}\rp+\mathcal{O}\left(\frac{|F|^4}{|m|^8}\right).
\end{equation}

Thus, to leading nontrivial order, we have reproduced \eqref{GauginoMass} for the gaugino masses and \eqref{ArResult},\eqref{SfermionBad} for the sfermion masses.

The scalar trilinears are produced by a 1-loop supergraph in the effective theory. If we denote the trilinear superpotential couplings by $W = \lambda_{ijk}\Phi_i\Phi_j\Phi_k$, then the coeffient of the corresponding scalar trilinear, $\phi_i\phi_j\phi_k$, is
\begin{equation}
  c_{ijk}=- 2\lambda_{ijk} \sum_{r=1}^3 g_r^4 \left(c_2(f_i,r)+c_2(f_j,r)+c_2(f_k,r)\right)
   \int \frac{d^4p}{(2\pi)^4} \frac{m \wt{B}_{1/2}^{(r)}(p^2)}{p^2(p^2 + |g_r^2 m \wt{B}_{1/2}^{(r)}(p^2)|^2)}
\end{equation}
Since $\wt{B}_{1/2}(p^2)$ vanishes at large momentum, the integral is convergent in the UV. In the IR, the integral is cut off at the gaugino mass, $m_\lambda^{(r)}=g_r^2m\wt{B}_{1/2}^{(r)}(0)$. For the Wilsonian coupling, $c_{ijk}(\mu)$, we should instead impose an IR cutoff at the scale, $\mu$ (and evaluate the gauge couplings, $g_r(\mu)$, at that scale). This incorporates the leading log corrections from running down from the scale $m$ to $\mu$. It is possible that $\wt{B}_{1/2}(p^2)$ is tiny, at the messenger scale, $p^2=m^2$ (subject to the constraints imposed by \eqref{B1result}). When that is the case, $c_{ijk}(m)\sim 0$, and the dominant contribution to the scalar trilinears come from the running down to lower energies.

Finally, we note that our results also show that there are no relations at higher order, at least not without invoking additional assumptions.  For instance, by adjusting $c_6'$ we can get any order $|F|^4$ contribution to the sfermion masses that we like.  Similarly, the order $F|F|^2$ contribution to $\wt{B}_{1/2}$ is unfixed, as are the two other linearly independent combinations of $\wt{C}_s$ at order $|F|^2$.  For higher orders in $F$ we can simply take the terms above multiplied by factors of $D^2M \ol{D}^2\ol{M}$ to show that all those terms are also independent.  Only the relations (\ref{BRelation}) and (\ref{CRelation}) are general.

\section{Multiple Spurions}\label{multispurion}
In the above analysis, we assumed that the effect of supersymmetry-breaking, as felt by the visible sector, could be summarized by the VEV of a single spurion $M= m + \theta^2 F$. More generally, we might have several spurions,
\begin{equation}
  M^{(r)} = m + \theta^2 F^{(r)}
\end{equation}
Since we have parametrized everything in terms of arbitrary functions, $f^{(r)}(p^2/|m|^2)$, we can, without loss of generality, take the lowest component to be the same for each gauge group factor. The global symmetry $G\supset SU(3)\times SU(2)\times U(1)$ of the hidden sector theory ensures that, within each gauge group factor, there is just a single spurion. In this case, the results for the soft supersymmetry breaking parameters are modified in the obvious way. The expression \eqref{GauginoMass}, for the gaugino masses, now reads
\begin{equation}\label{GauginoMassGen}                                                                
 m_\lambda^{(r)} = g_r^2 m \tilde{B}_{1/2}^{(r)}(0) = g_r^2 c^{(r)}  \frac{F^{(r)}}{m} +           
O\left(g_r^2\tfrac{F}{m}\tfrac{|F|^2}{|m|^4}\right)                                                                       
\end{equation}                                                                  
and the expression \eqref{ArResult} is modified, so that the sfermion masses are
\begin{equation}\label{SfermionMassGen}
m^2_{\tilde{f}} = \sum_{r=1}^3 g_r^4 c_2(f,r) \frac{c^{(r)}}{8\pi^2}\frac{|F^{(r)}|^2}{|M|^2} + O\left(\tfrac{|F|^4}{|M|^6}\right)
\end{equation}
While these spoil some of the detailed predictions of gauge mediation (like gaugino mass unification), it is comforting to see that \eqref{SfermionMassGen} is manifestly positive, unlike \eqref{SfermionBad},\eqref{ArBad}.

Having stated this, more general result, it should be noted that the most natural way to achieve that $c^{(r)}\equiv c$, for each of the factors in the Standard Model gauge group, is to posit that $G\supset SU(5)$, and that we gauge an $SU(3)\times SU(2)\times U(1)$ subgroup of this $SU(5)$. In that case, the same symmetry that guarantees gauge coupling unification also guarantees that each of the $F^{(r)}\equiv F$, reducing to the single spurion case analyzed above.

\section{D-term SUSY breaking}
\label{DTerms}

We would now like to apply our techniques to analyze a situation in which supersymmetry is broken in the hidden sector by a D-term.  That is, we imagine a $U(1)$ gauge group in the hidden sector whose vector superfield develops an auxiliary component VEV, $\langle D\rangle=d$.  This expectation value is real, has mass dimension two and carries no R-charge.  Gauge invariance demands that $d$ appear in the action as a field strength, $w^\al=\t^\al d$, or $\ol{w}^{\dot\al}=\ol{\t}^{\dot\al}d$, or supercovariant derivatives thereof.  After integrating out the hidden sector, the spurions $w^\al$ and $\ol{w}^{\dot\al}$ may enter the effective lagrangian which we will integrate over superspace.

Suppose that no hidden sector F-terms acquire VEVs, so that this D-term is the only source of SUSY breaking.  Let us apply our approach to this case, and identify all the terms that we can write down in the effective action that are built out of two MSSM gauge superfields, some number of spurions $w^\al$, $\ol{w}^{\dot\al}$, and the mass scale $m$.  Note first that since no fields with R-charge develop VEVs, the R-symmetry remains exact.  Hence the function $\wt{B}_{1/2}(p^2/|m|^2)$ must vanish, and thus the gaugino masses and A-terms must also vanish.  Note also that by Furry's theorem we must have an even number of vector superfields (including the spurions) appearing.  Proceeding as in \S\ref{HigherDerivativeTerms} we can enumerate the independent terms at order $d^2$,
\be
\begin{split}
J_1 =& |m|^{-6} W^\al f\left(\frac{\square}{|m|^2}\right)\s^\m_{\beta\dot\al}\p_\m D_\al W^\beta\, \ol{w}^{\dot\al}\lp D^\g w_\g\rp,\\
J_2 =& |m|^{-4}W^\al f\left(\frac{\square}{|m|^2}\right)W_\al\,\ol{w}_{\dot\al}\ol{w}^{\dot\al},\\
J_3 =& |m|^{-6}W^\al f\left(\frac{\square}{|m|^2}\right)\s^\m_{\al\dot\al}\p_\m\ol{W}^{\dot\al}\, \lp D^\beta w_\beta\rp\lp D^\g w_\g\rp,\\
J_4 =& |m|^{-4}W^\al f\left(\frac{\square}{|m|^2}\right)\ol{W}_{\dot\al}\, w_\al\ol{w}^{\dot\al}.
\end{split}
\ee
These are in addition to the supersymmetric terms, and can be added completely independently.  It is easy to check that each of the terms $J_1$, $J_2$, and $J_3$ contribute equal amounts to $\wt{C}_0$, $\wt{C}_{1/2}$, and $\wt{C}_1$, and so in particular do not contribute to the combination $\wt{C}_0-4\wt{C}_{1/2}+3\wt{C}_1$ which enters the sfermion masses.  Meanwhile, $J_4$ contributes an amount $2d^2|m|^{-4}f(p^2/|m|^2)$ to $\wt{C}_0$ and an amount $\hlf d^2|m|^{-4}f(p^2/|m|^2)$ to $\wt{C}_{1/2}$ (and nothing to $\wt{C}_1$), so it too does not contribute to the linear combination above.  This shows that the sfermion masses do not get a contribution at order $d^2/|m|^2$, though we can get contributions to both of the other independent linear combinations of the $\wt{C}_s$.

At order $d^4$, it is easy to see that we can get independent contributions to all the $\wt{C}_s$.  By multiplying the terms above by $|m|^{-4}(D^\al w_\al)(D^\beta w_\beta)=d^2/|m|^4$ we can get the two combinations already found, and by including a term
\be
|m|^{-8}W^\al f\left(\frac{\square}{|m|^2}\right)\s^\m_{\al\dot\al}\p_\m\ol{W}^{\dot\al}\,w^\beta w_\beta\,\ol{w}_{\dot\beta}\ol{w}^{\dot\beta},
\ee
we can get any contribution to $\wt{C}_{1/2}$ and hence to the sfermion masses.  Note that in this context there is nothing to guarantee that the resulting sfermion masses squared are positive.

Finally, we can consider the more realistic situation with both F-term and D-term SUSY breaking.  We assume that both $|F/m^2|$ and $d/m^2$ are small, though not necessarily the same order.  Then the lowest order contribution\footnote{Please recall that all of this discussion is at fixed lowest order in the standard model gauge couplings, and is only to all orders in the supersymmetry breaking parameters.} to $\wt{B}_{1/2}$, and hence also to the gaugino masses and A-terms, is the order $F$ contribution (\ref{B1result}).  This will receive corrections at all orders in both $|F|^2/|m|^4$ and $d^2/|m|^4$.  The leading contributions to the sfermion masses are of order $|F|^2/|m|^2$ contribution (\ref{ArResult}) and the above contribution, of order $d^4/|m|^6$. Again, both of these receive further corrections, which are a power series in $|F|^2/|m|^8$ and $d^2/|m|^4$.  If $d^2\gtrsim|Fm^2|$, then the $D$-term contributions to the sfermion masses squared can dominate. Since these can be of either sign, we are plunged back into the ``generic" situation of \cite{Meade:2008wd}. In particular, the sfermion masses squared can, in principle, be of either sign. Note, however, that we retain the leading predictions for the gaugino masses and A-terms, as these come from the F-term breaking discussed above.

\section*{Acknowledgements}
The research of the authors is based upon work supported by the National Science Foundation under Grant No. PHY-0455649.  D.~R. would like to thank the Aspen Center for Physics and J.~D.~would like to thank the Simons Workshop at SUNY Stonybrook for hospitality while the final stages of this paper were completed. J.~D.~would also like to thank D.~Berenstein, M.~Rocek and S.~Yost for very helpful discussions.

\vfill\eject
\appendix

\section{Explicit Treatment of Leading Order Higher Derivative Terms}
\label{LOTerms}

In this appendix we will focus on the terms of order $|M|^{-2}$.  We will list all possible index structures first, giving twenty-four different terms.  We will then use some identities to reduce the number of independent terms, and finally we will note that certain combinations of terms are total derivatives and hence will vanish when integrated over.  We won't deal with reality conditions until the very end.

First a list of terms.
\begin{equation}
\begin{aligned}
L_1 =& \left|M\right|^{-2} W^\al D^2 W_\al,\\
L_2 =& \left|M\right|^{-2}\lp D^\al W_\al\rp\lp D^\beta W_\beta\rp,\\
L_3 =& \left|M\right|^{-2}\lp D^\al W^\beta\rp\lp D_\al W_\beta\rp,\\
L_4 =& \left|M\right|^{-2}\lp D^\al W^\beta\rp\lp D_\beta W_\al\rp,\\
L_5 =& \left|M\right|^{-4}\lp\ol{M}D^\al M\rp\lp W_\al D^\beta W_\beta\rp,\\
L_6 =& \left|M\right|^{-4}\lp\ol{M}D^\al M\rp\lp W^\beta D_\al W_\beta\rp,\\
L_7 =& \left|M\right|^{-4}\lp\ol{M}D^\al M\rp\lp W^\beta D_\beta W_\al\rp,\\
L_8 =& \left|M\right|^{-4}\lp\ol{M}D^2M\rp\lp W^\al W_\al\rp,\\
L_9 =& \left|M\right|^{-6}\lp\ol{M}D^\al M\rp\lp\ol{M} D_\al M\rp\lp W^\beta W_\beta\rp,\\
L_{10} =& i\left|M\right|^{-2}\s^\m_{\al\dot\al}\lp\p_\m W^\al\rp\ol{W}^{\dot\al},\\
L_{11} =& i\left|M\right|^{-2}\s^\m_{\al\dot\al}W^\al\lp\p_\m\ol{W}^{\dot\al}\rp,\\
L_{12} =& \left|M\right|^{-2}\lp D^\al W_\al\rp\lp\ol{D}_{\dot\al}\ol{W}^{\dot\al}\rp,\\
L_{13} =& \left|M\right|^{-4}\lp\ol{M}D^\al M\rp W_\al\lp\ol{D}_{\dot\al}\ol{W}^{\dot\al}\rp,
\end{aligned}\quad
\begin{aligned}
L_{14} =& \left|M\right|^{-4}\lp M\ol{D}_{\dot\al}\ol{M}\rp\lp D^\al W_\al\rp\ol{W}^{\dot\al},\\
L_{15} =& \left|M\right|^{-4}\lp D^\al M\rp\lp\ol{D}_{\dot\al}\ol{M}\rp\lp W_\al\ol{W}^{\dot\al}\rp,\\
L_{16} =& \left|M\right|^{-2}\ol{W}_{\dot\al}\ol{D}^2\ol{W}^{\dot\al},\\
L_{17} =& \left|M\right|^{-2}\lp\ol{D}_{\dot\al}\ol{W}^{\dot\al}\rp\lp\ol{D}_{\dot\beta}\ol{W}^{\dot\beta}\rp,\\
L_{18} =& \left|M\right|^{-2}\lp\ol{D}_{\dot\al}\ol{W}_{\dot\beta}\rp\lp\ol{D}^{\dot\al}\ol{W}^{\dot\beta}\rp,\\
L_{19} =& \left|M\right|^{-2}\lp\ol{D}_{\dot\al}\ol{W}_{\dot\beta}\rp\lp\ol{D}^{\dot\beta}\ol{W}^{\dot\al}\rp,\\
L_{20} =& \left|M\right|^{-4}\lp M\ol{D}_{\dot\al}\ol{M}\rp\lp\ol{W}^{\dot\al}\ol{D}_{\dot\beta}\ol{W}^{\dot\beta}\rp,\\
L_{21} =& \left|M\right|^{-4}\lp M\ol{D}_{\dot\al}\ol{M}\rp\lp\ol{W}_{\dot\beta}\ol{D}^{\dot\al}\ol{W}^{\dot\beta}\rp,\\
L_{22} =& \left|M\right|^{-4}\lp M\ol{D}_{\dot\al}\ol{M}\rp\lp\ol{W}_{\dot\beta}\ol{D}^{\dot\beta}\ol{W}^{\dot\al}\rp,\\
L_{23} =& \left|M\right|^{-4}\lp M\ol{D}^2\ol{M}\rp\lp\ol{W}_{\dot\al}\ol{W}^{\dot\al}\rp,\\
L_{24} =& \left|M\right|^{-6}\lp M\ol{D}_{\dot\al}\ol{M}\rp\lp M\ol{D}^{\dot\al}\ol{M}\rp\lp\ol{W}_{\dot\beta}\ol{W}^{\dot\beta}\rp.
\end{aligned}
\end{equation}

Let us now try to tame this multitude.  First note that using (\ref{EpsilonIdentity}), we can derive the following relations
\begin{equation}
L_2+L_4=L_3,\qquad L_5+L_7=L_6,\qquad L_{17}+L_{19}=L_{18},\qquad L_{20}+L_{22}=L_{21}.
\end{equation}

Similarly, using (\ref{DWIdentity}) and (\ref{D^2WIdentity}), we find
\begin{equation}
L_1=4L_{11},\qquad L_{16}=-4L_{10},\qquad L_2=L_{12}=L_{17},\qquad L_5=L_{13},\qquad L_{20}=L_{14}.
\end{equation}

We still have fourteen independent terms.  Let us now determine which combinations are total derivatives.

\begin{equation}
\begin{split}
D^\al\lp\left|M\right|^{-2}W_\al D^\beta W_\beta\rp \Lra& -L_5+L_2-\hlf L_1=0,\\
D^\al\lp\left|M\right|^{-2}W^\beta D_\al W_\beta\rp \Lra& -L_6+L_3-L_1=0,\\
D^\al\lp\left|M\right|^{-2}W^\beta D_\beta W_\al\rp \Lra& -L_7+L_4-\hlf L_1=0,\\
D^\al\lp\left|M\right|^{-4}\ol{M}D_\al MW^\beta W_\beta\rp \Lra& -2L_9+L_8-2L_6=0,\\
D^\al\lp\left|M\right|^{-4}\ol{M}D^\beta MW_\al W_\beta\rp \Lra& -L_9+\hlf L_8-L_5-L_7=0,\\
\p_\m\lp\left|M\right|^{-2}\s^\m_{\al\dot\al}W^\al\ol{W}^{\dot\al}\rp \Lra& L_{10}+L_{11}=0,\\
D^\al\lp\left|M\right|^{-2}W_\al\ol{D}_{\dot\al}\ol{W}^{\dot\al}\rp \Lra& -L_{13}+L_{12}-2L_{11}=0,\\
D^\al\lp\left|M\right|^{-4}M\ol{D}_{\dot\al}\ol{M}W_\al\ol{W}^{\dot\al}\rp \Lra& -L_{15}-L_{14}=0,\\
\ol{D}_{\dot\al}\lp\left|M\right|^{-2}D^\al W_\al\ol{W}^{\dot\al}\rp \Lra& -L_{14}+2L_{10}+L_{12}=0,\\
\ol{D}_{\dot\al}\lp\left|M\right|^{-4}\ol{M}D^\al MW_\al\ol{W}^{\dot\al}\rp \Lra& L_{15}+L_{13}=0,\\
\ol{D}_{\dot\al}\lp\left|M\right|^{-2}\ol{W}^{\dot\al}\ol{D}_{\dot\beta}\ol{W}^{\dot\beta}\rp \Lra& -L_{20}+L_{17}-\hlf L_{16}=0,\\
\ol{D}_{\dot\al}\lp\left|M\right|^{-2}\ol{W}_{\dot\beta}\ol{D}^{\dot\al}\ol{W}^{\dot\beta}\rp \Lra& -L_{21}+L_{18}-L_{16}=0,\\
\ol{D}_{\dot\al}\lp\left|M\right|^{-2}\ol{W}_{\dot\beta}\ol{D}^{\dot\beta}\ol{W}^{\dot\al}\rp \Lra& -L_{22}+L_{19}-\hlf L_{16}=0,\\
\ol{D}_{\dot\al}\lp\left|M\right|^{-4}M\ol{D}^{\dot\al}\ol{M}\ol{W}_{\dot\beta}\ol{W}^{\dot\beta}\rp \Lra& -2L_{24}+L_{23}-2L_{21}=0,\\
\ol{D}_{\dot\al}\lp\left|M\right|^{-4}M\ol{D}_{\dot\beta}\ol{M}\ol{W}^{\dot\al}\ol{W}^{\dot\beta}\rp \Lra& -L_{24}+\hlf L_{23}-L_{20}-L_{22}=0,
\end{split}
\end{equation}

Finally, one should impose reality conditions, but this will just relate $W$ with $\ol{W}$ in the obvious ways.  So with the above results one can check that a good basis for the independent terms consists of $L_5$, $L_6$, $L_9$, and $L_{11}$, and that these should be added with their hermitian conjugates.

\bibliographystyle{utphys}
\bibliography{gaugemediation}

\end{document}